\documentstyle[twoside]{article}
\def\greaterthansquiggle{\raise.3ex\hbox{$>$\kern-.75em\lower1ex\hbox{$\sim$}}}
\def\lessthansquiggle{\raise.3ex\hbox{$<$\kern-.75em\lower1ex\hbox{$\sim$}}}
\newcommand{\beq}{\begin{equation}}
\newcommand{\eeq}{\end{equation}}
\newcommand{\beqa}{\begin{eqnarray}}
\newcommand{\eeqa}{\end{eqnarray}}
\newcommand{\beqan}{\begin{eqnarray*}}
\newcommand{\eeqan}{\end{eqnarray*}}
\newcommand{\ba}{\begin{array}}
\newcommand{\ea}{\end{array}}
\newcommand{\no}{\nonumber}
\newcommand{\ra}{\rightarrow}

\newcommand{\ve}{\varepsilon}

\newcommand{\wh}{\widehat}

\newcommand{\A}{{\cal A}}

\def\nz{\ifmmode {I\hskip -3pt N} \else {\hbox {$I\hskip -3pt N$}}\fi}
\def\zz{\ifmmode {Z\hskip -4.8pt Z} \else
       {\hbox {$Z\hskip -4.8pt Z$}}\fi}
\def\qz{\ifmmode {Q\hskip -5.0pt\vrule height6.0pt depth 0pt
       \hskip 6pt} \else {\hbox
       {$Q\hskip -5.0pt\vrule height6.0pt depth 0pt\hskip 6pt$}}\fi}
\def\rz{\ifmmode {I\hskip -3pt R} \else {\hbox {$I\hskip -3pt R$}}\fi}
\def\cz{\ifmmode {C\hskip -4.8pt\vrule height5.8pt\hskip 6.3pt} \else
       {\hbox {$C\hskip -4.8pt\vrule height5.8pt\hskip 6.3pt$}}\fi}

\begin{document}
\begin{titlepage}
\vspace{2cm}
\begin{center}
{\Large \bf Anomalies in Quantum Field Theory: Dispersion Relations and
Differential Geometry*}\\[1cm]
R. A. Bertlmann
\\
Institut f\"ur Theoretische Physik \\
Universit\"at Wien
\vfill
{\bf Abstract}\\
\end{center}
We present two different aspects of the anomalies in quantum field theory.
One is the dispersion relation aspect, the other is differential
geometry where we derive the Stora--Zumino chain of descent equations.
\vfill
\noindent *) Lecture given at the conference "QCD 94", Montpellier, France. \\
Supported by Fonds zur F\"orderung der wissen\-schaft\-lichen
Forschung, Project No. P8444--TEC.

\end{titlepage}

\section{INTRODUCTION}
Anomalies represent the breakdown of a classical symmetry, the breakdown
of the corresponding conservation law --- here the conservation of the axial
or chiral current --- by quantum effects [1,2]. They play an important
r\^ole in physics. Some recent work on the anomaly has been presented
already on this conference, for instance, the decay
$K^+ \ra \pi^+ \pi^0 \gamma$ [3], the spin of the proton [4,5], or the
$U(1)$--problem [6].

I want to concentrate more on the theoretical aspects. Anomalies have been
first discovered in perturbation theory by UV--regularizing a divergent
diagram [1,2]. But they are not just a regularization effect, they also
show up in quite different procedures. For example, in the method of
dispersion relations where they occur as an IR--singularity of the
transition amplitude [7,8], or within sum rules [9]. Using a quite
different approach, working with path integrals, the anomalies are
detected by the chiral transformation of the path integral measure [10].
In the last years a development in modern mathematical techniques
attracted much attention in describing the anomalies. This was differential
geometry [11--17], cohomology [18,19]and topology (Atiyah--Singer index
theorem) [20--29].

I would like to report on two aspects. One is the dispersion relation
aspect which is a conventional and practical technique, and the other is
the differential geometric approach, more precisely, the Stora--Zumino
chain of descent equations which has the appeal of mathematical elegancy,
and which provides an interesting relation among several `anomalous'
terms.

\section{DISPERSION RELATION \newline APPROACH}
The use of dispersion relations to calculate the axial anomaly has been
proposed originally by Dolgov and Zakharov [7] (for a review see
Ho\v rej\v s\'{\i}
 [30] and Kerson--Huang [31]). Even before, Kummer [32]
determined the anomaly in a dispersive way relying on an at that time
fashionable pion--nucleon model. We, however, want to demonstrate the
dispersion relation approach in a 2--dimensional example. The
advantage is its simplicity and we find all features of that method.
We follow here closely a work of Adam, Bertlmann and Hofer [8,33].

In quantum field theory we work with Green functions. They have to satisfy
the Ward identities (WI) --- the field theory equivalents of the classical
conservation laws --- to achieve the renormalizability of the theory.
In order to detect the anomaly in two dimensions we consider the
2--point function
\beq
\langle 0| T j_\mu(x) j^5_\nu(y)|0\rangle.
\eeq
Then we find for the axial Ward identity (AWI, and FT stands for Fourier
Transformation)
\beqa
q^\nu T^5_{\mu\nu} &=& \mbox{FT } \partial^\nu_y
\langle 0|T j_\mu(x) j^5_\nu(y)|0\rangle \no \\
&=& \mbox{FT } \langle 0|T j_\mu(x) \partial^\nu_y j^5_\nu(y)|0\rangle
\no \\
&=& 2m \mbox{ FT } i \langle 0|T j_\mu(x) P(y)|0\rangle \no \\
&=& 2m \; P^5_\mu
\eeqa
if we insert the classical result
\beq
\partial^\nu_y j^5_\nu(y) = 2im P(y) = 2im \bar \psi(y) \gamma_5
\psi(y).
\eeq
Analogously when considering the 2--point function
\beq
\langle 0|T j_\mu(x) j_\nu(y)|0\rangle
\eeq
we would get for the vector Ward identiy (VWI)
\beqa
q^\mu T_{\mu\nu} &=& \mbox{FT } \langle 0|T \partial^\mu_x j_\mu(x)
j_\nu(x) |0 \rangle = 0 \no \\
q^\nu T_{\mu\nu} &=& 0
\eeqa
if we rely on the classical conservation law
\beq
\partial^\mu_x j_\mu (x) = 0.
\eeq
But now we will calculate the Green functions explicitly, and we do it
in a dispersive way, and we check whether these WI's are satisfied.

In two dimensions we have
\beq
\gamma_\nu \gamma_5 = \ve_{\nu\lambda} \gamma^\lambda
\eeq
which implies for the amplitudes
\beq
T^5_{\mu\nu} = \ve_{\nu\lambda} T_\mu{}^\lambda
\eeq
and in addition we introduce the tensor structure
\beq
P^5_\mu (q) = \ve_{\mu\nu} q^\nu P(q^2).
\eeq
So it is enough to consider just the amplitude $T_{\mu\nu}$.
Let us investigate its structure. The general Lorentz decomposition gives
\beq
T_{\mu\nu}(q) = q_\mu q_\nu T_1(q^2) - g_{\mu\nu} T_2(q^2).
\eeq
All we require is that it vanishes at large $q^2$
\beq
T_{\mu\nu}(q) \stackrel{q^2 \ra \infty}{\longrightarrow} 0
\eeq
supplying
\beqa
T_1(q^2) &\stackrel{q^2 \ra \infty}{\longrightarrow}& 0 \no \\
T_2(q^2) &\stackrel{q^2 \ra \infty}{\longrightarrow}& \left\{ \ba{l}
0 \\ \mbox{const.} \neq 0 \ea \right. .
\eeqa
Then Cauchy's theorem provides the dispersion relations (DR) for the
amplitudes $T_1(q^2)$ and $T_2(q^2)$ (the amplitudes are analytic
except a cut starting at $t = 4m^2$). The unique choice for the amplitude
$T_1(q^2)$ is an unsubtracted DR (similarly for $P(q^2)$)
\beqa
T_1(q^2) &=& \frac{1}{\pi} \int_{4m^2}^\infty \frac{dt}{t-q^2}
\mbox{ Im } T_1(t) \no \\
P(q^2) &=& \frac{1}{\pi} \int_{4m^2}^\infty \frac{dt}{t-q^2}
\mbox{ Im } P(t).
\eeqa
Concerning $T_2(q^2)$, however, we have the option not to subtract the
DR or to subtract once. The natural choice would be again an unsubtracted
DR
\beq
T_2(q^2) = \frac{1}{\pi} \int_{4m^2}^\infty \frac{dt}{t-q^2}
\mbox{ Im } T_2(t)
\eeq
since the integral exists. Then the AWI is satisfied
\beqa
q^\nu T^5_{\mu\nu}(q) &=& q^\nu \ve_{\nu\lambda}(q_\mu q^\lambda T_1(q^2)
- g_\mu{}^\lambda T_2(q^2)) \no \\
&=& \ve_{\mu\nu} q^\nu T_2(q^2) \no \\
&=& \ve_{\mu\nu} q^\nu 2m P(q^2) \no \\
&=& 2m P^5_\mu(q)
\eeqa
since the imaginary parts of the amplitudes do fulfil the WI's in any case
\beqa
\mbox{AWI: } \mbox{Im } T_2(t) &=& 2m \mbox{ Im } P(t) \no \\
\mbox{VWI: } \mbox{Im } T_2(t) &=& t \mbox{ Im } T_1(t) .
\eeqa
However, considering the VWI we find
\beqa
q^\mu T_{\mu\nu} &=& q_\nu (q^2 T_1(q^2) - T_2(q^2)) \no \\
&=& q_\nu \;\frac{1}{\pi} \int_{4m^2}^\infty
\frac{dt(q^2 - t)}{t - q^2} \mbox{ Im } T_1(t) \no \\
&=& q_\nu \left( - \frac{1}{\pi} \right) \int_{4m^2}^\infty
\mbox{Im } T_1(t) \no \\
&=& \A \; q_\nu
\eeqa
with the anomaly
\beq
\A = - \frac{1}{\pi} .
\eeq
In the last step we had to do an explicit calculation, we calculated the
imaginary part of a fermion loop with help of the Cutkosky rules [8,33]
\beq
t \mbox{ Im } T_1(t) = \frac{2m^2}{t} \left( 1 - \frac{4m^2}{t}
\right)^{-1/2} .
\eeq
Now, for physical reasons, in order to restore gauge invariance, to satisfy
VWI, we subtract the DR for
\beqa
T_2^R(q^2) &=& T_2(q^2) - T_2(0) \no \\
&=& \frac{q^2}{\pi} \int_{4m^2}^\infty \frac{dt}{t(t - q^2)}
\mbox{ Im } T_2(t)
\eeqa
then the subtraction constant represents the ano\-maly
\beqa
- T_2(0) &=& - \frac{1}{\pi} \int_{4m^2}^\infty \frac{dt}{t}
\mbox{ Im } T_2(t) \no \\
&=& - \frac{1}{\pi} \int_{4m^2}^\infty dt \mbox{ Im } T_1(t) \no \\
&=& - \frac{1}{\pi}= \A .
\eeqa
Furthermore we have
\beqa
T_2^R(q^2) &=& \frac{q^2}{\pi} \int_{4m^2}^\infty
\frac{dt}{t(t-q^2)} \mbox{ Im } T_2(t) \no \\
 &=& \frac{q^2}{\pi} \int_{4m^2}^\infty
\frac{dt}{t-q^2} \mbox{ Im } T_1(t) \no \\
&=& q^2 T_1(q^2)
\eeqa
and we achieve the familiar gauge invariant tensor structure (with
$T_1 \equiv T$)
\beq
T_{\mu\nu}(q) = (q_\mu q_\nu - q^2 g_{\mu\nu}) T(q^2).
\eeq
The AWI, on the other hand, contains the ano\-maly
\beqa
q^\nu T^5_{\mu\nu} &=& q^\nu \ve_{\nu\lambda} (q_\mu q^\lambda T_1(q^2)
- g_\mu{}^\lambda T^R_2(q^2)) \no \\
&=& \ve_{\mu\nu} q^\nu (T_2(q^2) - T_2(0)) \no \\
&=& \ve_{\mu\nu} q^\nu (2m P(q^2) + \A).
\eeqa
So the axial current is not conserved but provides the well known
anomaly result (we take $m = 0$)
\beq
\partial^\nu j^5_\nu = \frac{e}{\pi} \ve_{\nu\mu} \partial^\nu A^\mu
= \frac{e}{2\pi} \ve_{\nu\mu} F^{\nu\mu} .
\eeq
If we subtract the amplitude $T_2(q^2)$ at some arbitrary point $q^2_a$
then the anomaly is distributed on both WI's, the AWI and the VWI.
There is no choice to get rid of the anomaly at all.

The source of the anomaly in this dispersive procedure is the existence
of the superconvergence sum rule
\beq
\int_{4m^2}^\infty dt \mbox{ Im } T_1(t) = 1.
\eeq

The anomaly corresponds to a threshold singularity of Im~$T_1(t)$ at
$t = 4m^2$ approaching a $\delta$--function for $m \ra 0$
\beqa
\lefteqn{\lim_{m \ra 0} \mbox{ Im } T_1(t) =} \no \\
&=& \lim_{m \ra 0} \frac{2m^2}{t^2}
\left( 1 - \frac{4m^2}{t}\right)^{-1/2} \theta(t - 4m^2) \no \\
&=& \delta(t).
\eeqa

\section{STORA--ZUMINO CHAIN OF \newline DESCENT EQUATIONS}
Now we return to four dimensions and to nonabelian gauge fields.
There exist two types of anomalies (for a review see Jackiw [34]),
the singlet anomaly and the nonabelian anomaly (Bardeen's result
[35]). In terms of differential forms the explicit expressions are:\\
singlet anomaly
\beqa
\A &=& d*j^5 = \frac{1}{4\pi^2}\mbox{ tr } FF \no \\
&=& \frac{1}{4\pi^2}
\mbox{ tr } d\left(AdA + \frac{2}{3} A^3 \right)
\eeqa
nonabelain anomaly
\beqa
- G^a[A] &=& (D*j)^a \no \\
&=& \frac{1}{24\pi^2} \mbox{ tr } T^a d\left(AdA + \frac{1}{2}A^3\right).
\eeqa
(Note that in the second case the field and the current are of
positive chirality, for negative chirality we get a sign change, the
chirality index is suppressed.)  Although the two expressions resemble
each other very closely they are of different origin. But there exists
a very peculiar relation between these two anomalies.
The nonabelian anomaly in $2n$ dimensions is related to the singlet
anomaly in $(2n + 2)$ dimensions via some chain of differential
geometric equations. This is the Stora--Zumino (SZ) chain of descent
equations we want to present now [11--16].

We carry out pure mathematics, the construction is actually algebraic.
Instead of the trace we use the more general, symmetric invariant
polynomial $P(F^n)$. Since $P(F^n)$ is closed (due to the Bianchi
identity $DF = 0$) it is locally exact (Poincar\'e lemma)
\beq
P(F^n) = d Q_{2n-1} (A,F).
\eeq
The polynomial $Q_{2n-1}$ of form degree $(2n-1)$ --- called
Chern--Simons form --- can be calculated explicitly
\beq
Q_{2n-1} (A,F) = n \int_0^1 dt \; P(A,F_t^{n-1})
\eeq
where
\beq
F_t = t F + (t^2 - t) A^2
\eeq
(geometrically, we consider a trivial fibre bundle).
Equation (30) -- (32), also called transgression formula, is the starting
point of the following mathematical operations.

Stora [12] noticed that a shift in both, in the gauge potential ---
geometrically the connection --- and in the derivative
\beqa
A &\ra& \wh A = A + v \no \\
d &\ra& \Delta = d + s
\eeqa
where $v$ represents the Faddeev--Popov ghost and $s$ the BRS operator,
leaves the field strength --- geometrically the curvature --- invariant
\beq
dA + A^2 = F(A) \equiv \wh F(\wh A) = \Delta \wh A + \wh A^2.
\eeq
Identity (34) holds by virtue of the familiar BRS equations and is named
by Stora ``Russian formula''.

Due to the ``Russian formula'' we can equate the transgression formula
(30) with its shifted version
\beqa
\lefteqn{\Delta Q_{2n-1}(A + v,F) =} \no \\
&=& P(\wh F^n) \equiv P(F^n) = d Q_{2n-1}(A,F).
\eeqa
Next we expand the Chern--Simons form in powers of the Faddeev--Popov ghost $v$
\beqa
\lefteqn{Q_{2n-1}(A+v,F) =} \no \\
&=& Q^0_{2n-1}(A,F) + Q^1_{2n-2}(v,A,F) \no \\
&& \mbox{} + Q^2_{2n-3}(v,A,F) + \ldots + Q^{2n-1}_0(v)
\eeqa
where the upper index denotes the powers of $v$ and the lower the form
degree. We insert this expansion into equation (35)
\beqa
\lefteqn{(d+s) Q^0_{2n-1} + (d+s) Q^1_{2n-2} + (d+s) Q^2_{2n-3} } \no \\
&&\mbox{} + \ldots + (d+s) Q^{2n-1}_0 = d Q^0_{2n-1},
\eeqa
we compare the terms of same form degree and same power in $v$ then we
obtain a chain of descent equations (we have followed here Stora's
approach [12], the view of Zumino [14] is more geometric but equivalent)
\beqa
P(F^n) - d Q^0_{2n-1} &=& 0 \no \\
s Q^0_{2n-1} + d Q^1_{2n-2} &=& 0 \no \\
s Q^1_{2n-2} + d Q^2_{2n-3} &=& 0 \no \\
\ldots && \no \\
s Q_1^{2n-2} + d Q_0^{2n-1} &=& 0 \no \\
s Q_0^{2n-1} &=& 0.
\eeqa
The chain we can solve now term by term. Starting with the Chern--Simons
form we apply the BRS operator $s$, respect the BRS rules and we find
$Q^1_{2n-2}$, and so forth. The explicit solutions in case of $n = 3$
are (they are not unique, general formulae are given by Zumino [16])
\beqa
Q^0_5 &=& \mbox{tr }\left[A(dA)^2 + \frac{3}{2} A^3 dA
+ \frac{3}{5} A^5\right] \no \\
Q^1_4 &=& \mbox{tr }vd \left(AdA + \frac{1}{2} A^3\right) \no \\
Q^2_3 &=& - \frac{1}{2} \mbox{ tr } [(v^2 A + vAv + Av^2)dA
+ v^2 A^3] \no \\
Q^3_2 &=& \frac{1}{2} \mbox{ tr } [- v^3 dA + A vAv^2] \no \\
Q^4_1 &=& \frac{1}{2} \mbox{ tr } v^4 A \no \\
Q^5_0 &=& \frac{1}{10} \mbox{ tr } v^5.
\eeqa
What we observe now is that $Q^1_4$ provides precisely the nonabelian
anomaly (29) besides the normalization. The reason is that equation
\beq
s Q^1_{2n-2} + d Q^2_{2n-3} = 0
\eeq
represents a local version of the Wess--Zumino consistency condition
--- the condition which defines the anomaly. So we can identify the
chain term $Q^1_{2n-2}$ (mathematics) with the anomaly $G(v,A)$
(physics)
\beq
G(v,A) = \int_M v^a G^a[A] = N \int_M Q^1_{2n-2} .
\eeq
The normalization $N$ we must take from somewhere else, from perturbation
theory or from path integral methods or from topological methods.

Let us summarize. On pure mathematical grounds we can derive a system
of equations --- the SZ chain of descent equations --- which we can
solve. Several terms have a correspondence in physics what we list at the
end.

Physical meaning of the chain terms:
\begin{itemize}
\item $P(F^n) = \mbox{tr }F^n$

singlet anomaly in $2n$ dimensions
\item $Q^0_{2n-1}$

Chern--Simons form, ingredient for topological field theoires
[36--39]

\item $Q^1_{2n-2}$

nonabelian anomaly, solution of the Wess--Zumino consistency condition
[12,14]

\item $Q^2_{2n-3}$

Schwinger term in an equal time commutator of Gauss--law operators [40]

\item $Q^3_{2n-4}$

represents the violation of the Jacobi identity for velocity operators
in the presence of a magnetic monopole [41,42]
\end{itemize}

\subsection*{Acknowledgement}
We would like to thank Stephan Narison, the organizer of the conference
``QCD 94'', for providing such a pleasant and stimulating atmosphere.

\end{document}